\documentclass[epj]{svjour}
\usepackage{graphicx}
\usepackage{epsfig}
\usepackage{amsmath,amssymb}
\usepackage[percent]{overpic}

\usepackage{color} 

\usepackage[normalem]{ulem}
\usepackage{cancel}

\begin{document}

\markboth{C. Cakan et al.}{Heterogeneous Delays in Neural Networks}
\graphicspath{{figs/}}

\title{Heterogeneous Delays in Neural Networks}

\author{Caglar Cakan, 
Judith Lehnert, 
Eckehard Sch\"oll}
\institute{Institut f{\"u}r Theoretische Physik, TU Berlin, Hardenbergstra\ss{}e 36, 10623 Berlin, Germany}
\mail {schoell@physik.tu-berlin.de}

\date{\today}
\abstract{
We investigate heterogeneous coupling delays in complex networks of
excitable elements described by the FitzHugh-Nagumo model. 
The effects of discrete as well as of uni- and
bimodal continuous distributions are studied with a focus on different
topologies, i.e., regular, small-world, and random networks. In the
case of two discrete delay times resonance effects play a major role:
Depending on the ratio of the delay times, various characteristic
spiking scenarios, such as coherent or asynchronous spiking,
arise. For continuous delay distributions different dynamical patterns emerge
depending on the width of the distribution. For small distribution
widths, we find highly synchronized spiking, while for intermediate
widths only spiking with low degree of synchrony persists, which is associated with traveling disruptions, partial amplitude death, or subnetwork synchronization, depending sensitively on the network topology. If the inhomogeneity of the coupling delays becomes too large, global amplitude death is induced. 
}

\maketitle

\section{Introduction}\label{sec:intro}
Over the past decades the study of networks has gained increasing
importance due to its widespread applicability such as in social sciences, computer science,
economics, biology, physiology, and ecology.
In the context of dynamics on networks, synchronization is a field of
high interest \cite{PIK01}. It is an important phenomenon, for instance, in neuroscience \cite{ROE97,POE01,ROS05,SIN07,VIC08,MAS08,LEH11,KAN11a,PER11a,KEA12}. The master stability
function is a powerful tool to investigate the stability of zero-lag,
cluster and group synchronization \cite{PEC98,KES07,SOR07,DAH12} but generally is
limited to networks of identical systems with one discrete delay
time. Only under the condition of commuting coupling matrices, two or
more discrete delay times can be considered \cite{SOR07,DAH12}. However, real-world networks are not limited to cases well
described by commuting coupling matrices and discrete delay times but
are often characterized by a complex topology, for example of random
or small-world type, and heterogeneous delays. In the brain, the
length, the diameter, and the kind of the axons between the neurons
determine the nerve conduction velocity. Therefore the propagation
speed between neurons can vary between $1$ and $100$ mm/ms
\cite{KOC99}. The aim of this paper is to investigate synchronization and other space-time patterns in the presence of heterogeneous delay times.
In particular we consider two discrete delay
times as well as unimodal and bimodal delay
distributions.   In the context of the brain, a bimodal distribution
is a good first approximation if coupling on two
different length scales is considered, i.e., nearby connections within
brain areas associated with short delays, and links between distant
areas characterized by long delays.

The paper is organized as follows: In Sec. 2 we introduce the neural model and the delay coupling. Section 3 discusses possible spiking scenarios in regular,
small-world, and random networks if the coupling delays are drawn from a
unimodal distributions. In Sec. 4, we explore the role of resonance
effects in the presence of two different discrete delay times. Section
5 extends the results of Secs. 2 and 3 by investigating a bimodal
delay distribution. A conclusion is given in Sec. 6.

\section{Model}\label{sec:model}
The local dynamics of each node in the network is modeled by the
FitzHugh-Nagumo differential equations \cite{FIT61,NAG62}. The
FitzHugh-Nagumo model is paradigmatic for excitable dynamics close
to a Hopf bifurcation \cite{LIN04}, which is not only characteristic
for neurons but also occurs in the context of other systems ranging
from electronic circuits \cite{HEI10} to cardiovascular tissues and
the climate system \cite{MUR93,IZH00a}.
Each node of the network is described as follows:
\begin{align}
\varepsilon \dot u_{i} &= u_{i} -\frac{u_{i}^3}{3}-v_{i} + C \sum_{j=1}^N G_{ij}
\left[u_j(t-\tau_{ij})-u_i(t)\right],  \nonumber\\ 
\dot v_i&=u_i+a,  \quad i=1,\ldots,N,                     \label{eq:FHN_Coupled}
\end{align}
where $u_i$ and $v_i$ denote the activator and inhibitor variable of
the nodes $i=1,\ldots,N$, respectively, and $\varepsilon$ is a time-scale
parameter and typically small (here we will use $\varepsilon=0.01$),
meaning that $u_i$ becomes a fast variable while $v_i$ changes
slowly. In the uncoupled system ($C=0$), $a$ is the threshold parameter: 
For $a>1$ the system is excitable while for $a<1$ it exhibits self-sustained periodic firing. This is due to a supercritical Hopf bifurcation at $a=1$ with a locally stable fixed point for $a>1$ and a stable limit cycle for $a<1$. We choose
$a=1.3$ such that the system operates in the  excitable regime. 

 The coupling matrix $\textbf{G}=\{G_{ij}\}$ defines which nodes are
connected to each other. An invariant synchronization manifold will
only exist if $\textbf{G}$ has a constant row sum; without loss of
generality we assume the row sum to be unity, i.e., $\sum_jG_{ij}=1$,
$i=1,\ldots,N$. We construct the matrix $\textbf{G}$ by setting the
entry $G_{ij}$ equal to 1 (0) if the $j$th node couples (does not
couple) into the $i$th node. After repeating this procedure for all
entries of $G_{ij}$, we normalize each row to unity. The overall coupling
strength is given by $C$. Throughout the paper, we use bidirectional
coupling, which means that signals can always be transmitted in both
directions. This makes $\textbf{G}$ a symmetric matrix (before we
normalize each row sum to $1$). $\textbf{T}=\{\tau_{ij}\}$ is the delay
matrix, i.e., $\tau_{ij}$ is the time the signal needs to propagate
from the $j$th to the $i$th node.

\section{Unimodal delay distributions in complex networks}\label{sec:unimodal}

The first step in going from one discrete delay time to more realistic
models with heterogeneous delay times is to consider a unimodal
distribution. In the following, we choose the elements of
the delay matrix $\textbf{T}$ randomly from a  normal distribution
$\mathcal{N}(\tau_{\mu},\sigma^2)$ with mean $\tau_{\mu}$ and standard
deviation $\sigma$.

 In Ref.~\cite{LEH11}, system~\eqref{eq:FHN_Coupled} with a $\delta$-distribution of the
delay times was investigated. For excitatory coupling -- i.e., all
entries of $\textbf{G}$ are positive -- it was shown that synchronized spiking with an inter-spike
interval (ISI) of $\tau_{\mu}$  is always stable independently of
coupling strength and delay time (as long as both are large enough to
induce any spiking at all). In this Section, we will discuss how robust
these results are if we increase $\sigma$. In particular we will
focus on the effect of the underlying topology -- regular, small-world, or random -- on the dynamics. These topologies are constructed as follows:
In a regular ring network each node is connected with equal strength
to its $k$ nearest neighbors to the left and to the right, i.e., the node
degree is $2k$. If additional excitatory links are added with a probability $p$ to such  a regular network, a small-world network arises
\cite{WAT98,MON99,NEW99b}. In a random network each node is linked
with probability $p$ to every other node \cite{RAP57,SOL51,ERD59,ERD60}.

Depending on the distribution width, the topology, and initial conditions there are essentially three different
types of dynamics observable: \textit{Highly-synchronous spiking},
\textit{spiking}, and \textit{global amplitude death}, from which the 
\textit{spiking} can be subdivided into several subclasses. 

\subsection{Spiking patterns and amplitude death}\label{sec:patterns}
\subsubsection*{Highly-synchronous spiking}
If $\sigma$ is small enough, stable synchronization with an ISI of
$\tau_{\mu}$ will persist, while the spikes will be broader than for a
delta distribution.

The quality of synchronization can be measured using the Kuramoto order parameter \cite{KUR84}:
 \begin{eqnarray}\label{eq:Kuramoto}
R=\frac{1}{N_s} \left|\sum_{j=1}^{N_s }  e^{i\phi_j}\right|,
\end{eqnarray}
where $N_s$ is the number of spiking nodes. In other words, we do not
include non-spiking nodes when calculating the Kuramoto order parameter.
 $\phi_j$ is the phase of the $j$th oscillator. $\phi_j$ can be
defined  as
\begin{eqnarray}\label{eq:phase}
\phi_j=2 \pi \frac{t-t_n}{t_n-t_{n-1}},
\end{eqnarray}
where $t_n$ is the time of the last spike of the  $j$th neurons \cite{ROS01}. For
$R=1$, perfect phase synchronization is reached; if $R\approx 0$,
the network is desynchronized. We consider a network
as \textit{highly synchronized} if $R>0.99$ after all transient effects have vanished.  

\subsubsection*{Spiking}
For intermediate $\sigma$,
different dynamical subclasses  can be
observed where the network still exhibits spikes, but not necessarily in a highly
synchronized manner: approximate synchronization, traveling disruptions, and partial amplitude death.

\begin{figure*}
\begin{center}
 \begin{overpic}[width=1\textwidth]{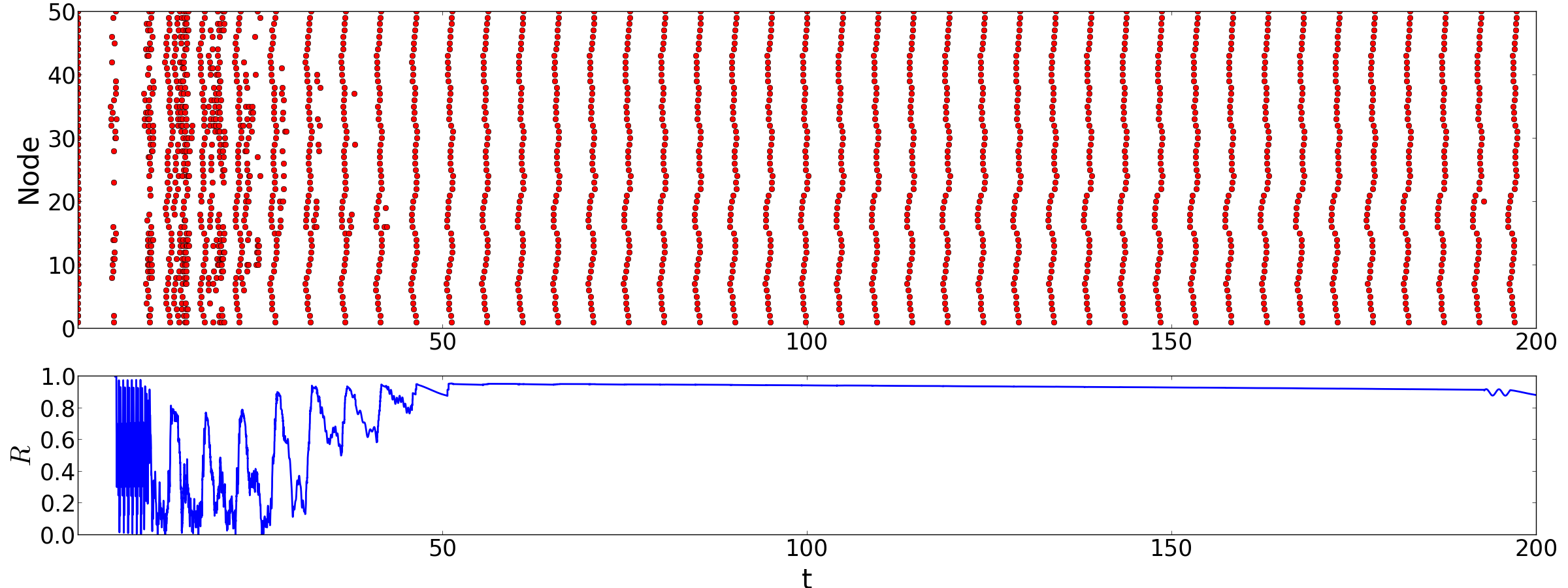}
  \put(0,36){(a)}
  \put(0,13){(b)}
\end{overpic}
\caption{(a) Approximately synchronized spiking pattern (red dots)  and (b) order parameter $R$ (blue line) vs time for a small-world network with a delay distribution with standard deviation $\sigma=0.2$.
Parameters: $N=50$, $p=0.51$, $k=2$, $C=1.0$, $a=1.3$,
$\varepsilon=0.01$ and $\tau_{\mu}=5$. The initial condition is
$u_i=-a$, $v_i=a-a^3/3$ for all $i=1,...,50$. The history function of
all nodes is the spiking state. }\label{fig:sync-commute}
\end{center}
\end{figure*}

The first scenario is that all nodes in the network still synchronize
but because of the non-zero width of the delay distribution the
incoming spikes do not arrive exactly at the same time but
with some slight deviations. Thus, $R$ drops below $0.99$,
i.e., the network is not any longer highly synchronized. Figure~\ref{fig:sync-commute}(a) shows as an example the time series of \textit{approximate synchronization}: The spike times are marked by red dots in panel (a) and the Kuramoto order parameter $R(t)$ is plotted in panel (b).  After some transienst time the spikes synchronize fast. However, $R$ remains below $0.99$. 

\begin{figure*}[h!t]
\begin{center}
\includegraphics[width=1\textwidth]{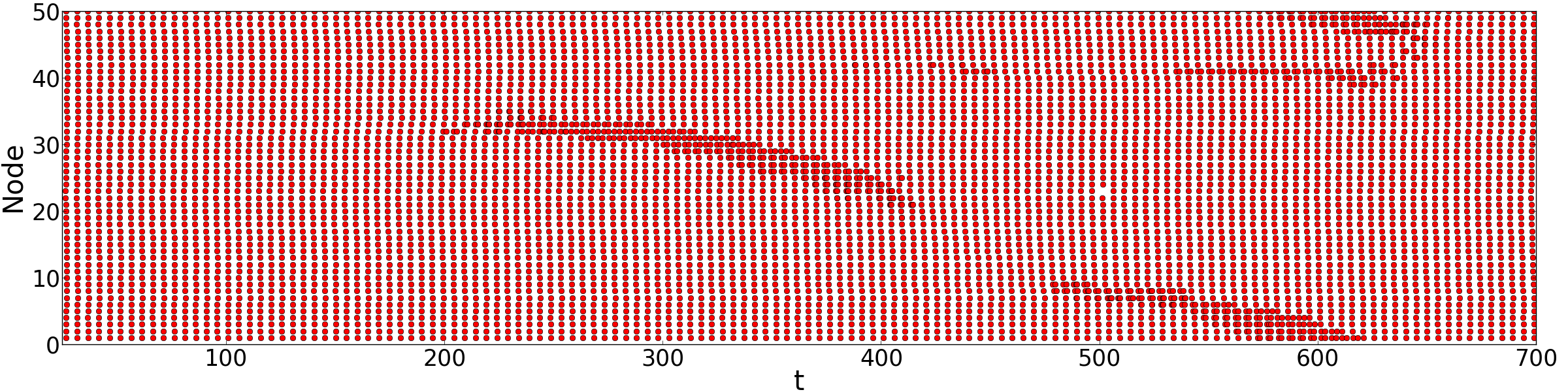}
\caption{Spiking patterns showing \textit{traveling disruptions} in a regular ring network with $N=50$
  nodes and $\sigma=0.2$. Red dots indicate spikes. Other parameters as in Fig.~\ref{fig:sync-commute}.
   }\label{fig:traveling-wave}
\end{center}
\end{figure*}

A fairly well synchronized behavior is the most common spiking pattern,
but in regular-ring structures as well as in some
small-world realizations a different type of behavior can be observed
as well: In Fig. \ref{fig:traveling-wave}(a) disruption travels along
many nodes in a ring network without causing amplitude death to the
network. Such \textit{traveling disruptions} can arise for fairly high standard deviations, e.g., in Fig. \ref{fig:traveling-wave}, $\sigma=0.2$, however, the
probability for this behavior is quite low since amplitude death  is much more common as will be discussed later.

\begin{figure*}
\begin{center}
\includegraphics[width=1\textwidth]{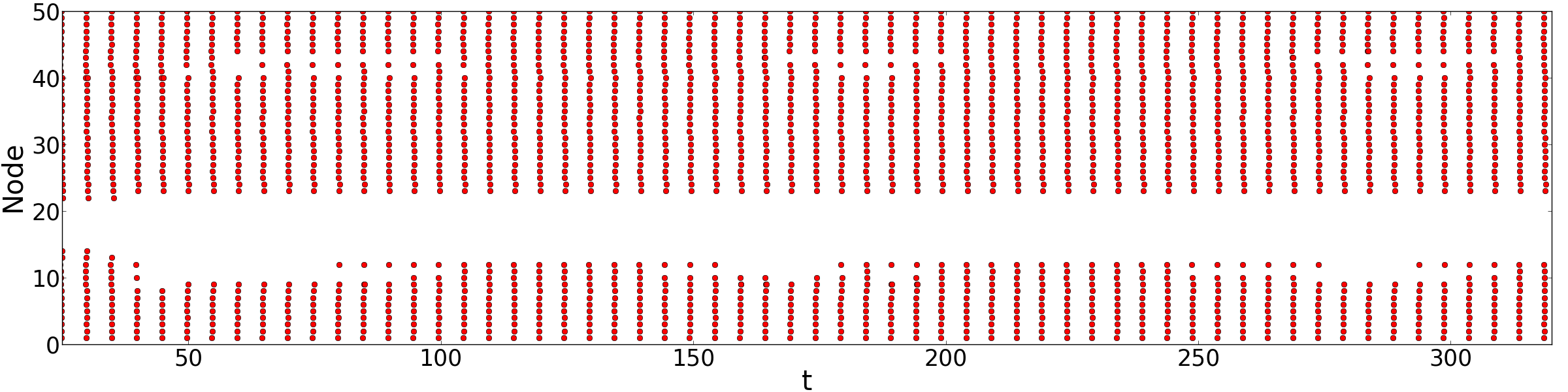}
\caption{Spiking pattern showing partial amplitude death in a
  small-world network ($N=50$, $p=0.51$) for a delay distribution with standard deviation of $\sigma=0.12$. Red dots mark
  spikes.  Other parameters as in Fig.~\ref{fig:sync-commute}.
  }\label{fig:nearly-dying}
\end{center}
\end{figure*}

Furthermore, networks can be observed where only a subset of nodes
spikes, while the other nodes undergo \textit{partial amplitude
  death}. This is the case if, by chance, in a fairly isolated subnetwork the deviation of delay times is smaller than the deviation in the whole network. In large random networks the probability for this is small, since fairly isolated subnetworks arise rarely as they require some
kind of ordered structure in the network.
In case of small-world networks
though, it is more likely that a subnetwork can maintain stably
synchronized spiking while other regions of the network undergo
amplitude death. This scenario is depicted in
Fig. \ref{fig:nearly-dying}. One can also observe cases where a part
of the network stops spiking temporarily but then gets excited again
and fires synchroneously with the rest of the network. This reanimation is also observable in Fig. \ref{fig:nearly-dying} around node no. 43.

\subsubsection*{Global Amplitude death}
If  $\sigma$ becomes too large, amplitude death is induced, i.e., 
all nodes remain in the fixed point and no spikes are released.

Amplitude death is caused by two different factors. First, each node needs excitation from many neighbors at the same time. As $\sigma$ increases, the probability that enough spikes from the neighboring neurons arrive sufficiently close in time decreases. Furthermore, the nodes' neighbors which spike too early or too late and  which are therefore already or still in the fixed point $(u_*,v_*)$  will pull the node back as they give rise to a negative coupling term of the form $u_*-u_i$ in Eq.~\eqref{eq:FHN_Coupled}, where we considered the effect on the $i$th node.   

Second, even if a spike can be excited by the arriving spikes, for
large $\sigma$  retarded spikes are likely to perturb the trajectory of the already spiking node. The result is a  spike with a low amplitude, which will be fed back into the network. If this happens on a global scale, i.e., if there is no fairly isolated subnetwork, which can maintain the amplitude of the spikes, the spikes will be damped in the course of time and will eventually lead to global amplitude death. 
\begin{figure}
\begin{center}
 \begin{overpic}[width=0.5\textwidth]{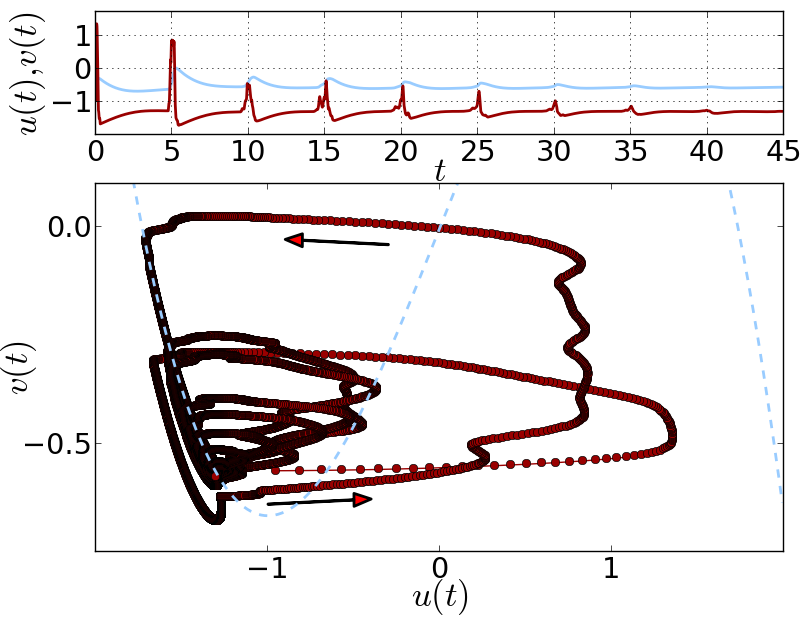}
  \put(12.5,72){(a)}
  \put(12.5,50){(b)}
\end{overpic}
\caption{(a) Time series of $u$ (red solid) and $v$ (blue dotted), and
  (b) phase space of a single node undergoing amplitude death. The
  node is part of a small-world network ($N=50$, $p=0.51$) with $\sigma=0.15$. (b) Trajectory (red solid) and $u$-nullcline (blue dashed). Other parameters as in Fig.~\ref{fig:sync-commute}.\label{fig:traject_ampdeath}}
\end{center}
\end{figure}

Both of the mentioned effects can be seen in the trajectory depicted
in Fig. \ref{fig:traject_ampdeath}. The first effect prevents the
trajectory to reach the far right nullcline (dashed blue) in the phase
space, which would have been reached in the absence of the negative force pulling it to the left. The second effect causes the trajectory to wiggle around on its round trip, instead of allowing a smooth course. This is due to excitations arriving during the round trip which pull the trajectory back and forth. Both effects lead to an decreasing amplitude and eventually to amplitude death.

\subsection{Statistical analysis}\label{ssec:statistics}
\begin{figure*}[ht!]
\begin{center}
 \begin{overpic}[width=1\textwidth]{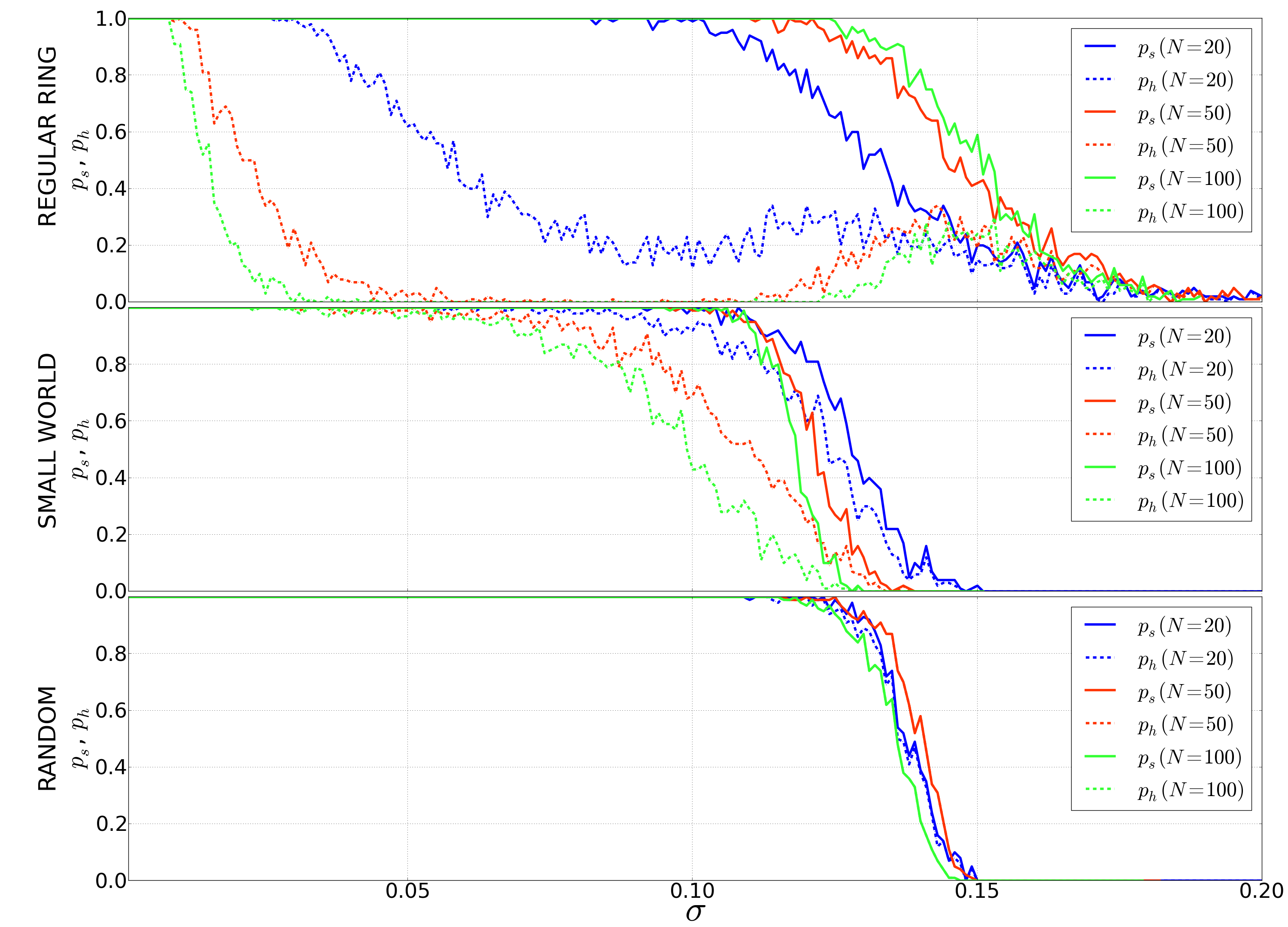}
  \put(11,68){(a)}
  \put(11,46){(b)}
  \put(11,23){(c)}
\end{overpic}
\caption{Probability of spiking $p_s$ (solid lines) and of highly
  synchronized spiking $p_h$ (dotted lines) vs standard deviation of
  delay distribution $\sigma$ for (a) a regular ring networks with
  $k=2$, i.e, each node is connected to its nearest and next-nearest
  neigbor to the left and to the right,   (b) small-world networks with
  $k=2$ and $p=0.51$ and (c) random networks with $p=0.51$ and different network sizes $N=20$ (blue), $N=50$ (red),  $N=100$ (green). Other parameters as in Fig.~\ref{fig:sync-commute}.}
\label{fig:stats}
\end{center}
\end{figure*}

Which kind of dynamics will take place on a network
depends on the topology, the width of the delay distribution, and on
initial conditions. For a systematic study, we calculate $p_s(\sigma)$ and $p_h(\sigma)$, where $p_s(\sigma)$ and
$p_h(\sigma)$ are the probabilities that for a given realization of
the delay matrix $\mathbf{T}$ with the standard deviation $\sigma$ and a given
realization of the network topology, the network shows any kind of
spiking behavior ($p_s$) or highly synchronized spiking ($p_h$),
respectively. Note that $p_h \le p_s$ as the highly synchronized networks are
a subset of the spiking ones. Figure~\ref{fig:stats} shows
$p_h(\sigma)$ (dotted lines) and $p_s(\sigma)$ (solid lines) for (a) a regular ring, (b) a small-world network, and (c) a random network.

\begin{figure*}[ht!]
\begin{center}
 \begin{overpic}[width=1\textwidth]{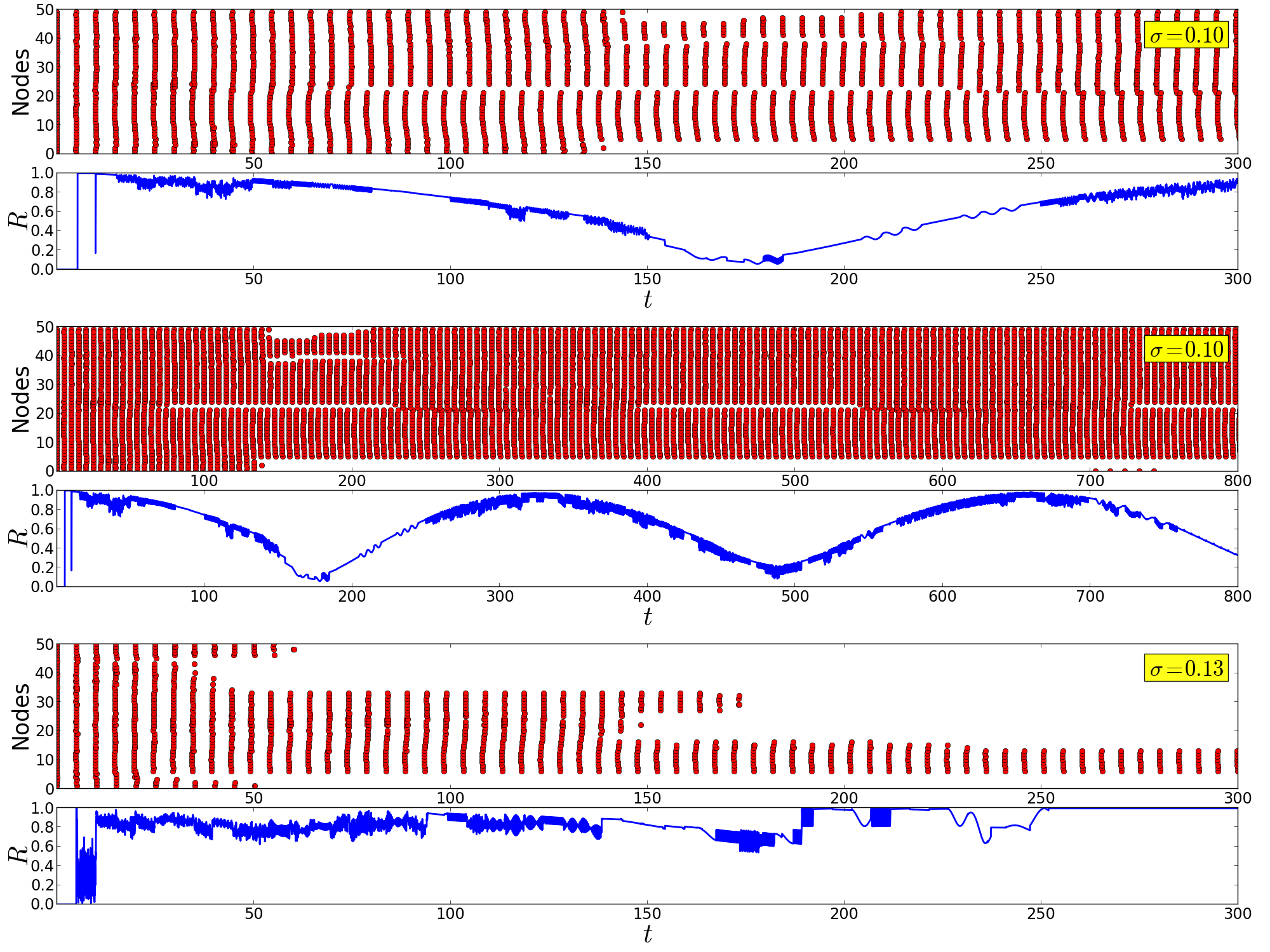}
  \put(0,74){(a)}
  \put(0,61.2){(b)}
  \put(0,49){(c)}
  \put(0,36.2){(d)}
  \put(0,24){(e)}
  \put(0,11.2){(f)}
\end{overpic}
\caption{Spiking patterns showing \textit{subnetwork synchronization} in a regular ring network with $N=50$ nodes for different $\sigma$: (a)-(d) $\sigma=0.1$; (e),(f)$ \sigma=0.13$. Panel (a),(c),(e): Red dots indicate spikes; Panel (b),(d),(f): Kuramoto order parameter $R$. Note that (c),(d) show the same simulation as (a),(b) for a longer time series. Other parameters as in Fig.~\ref{fig:sync-commute}.}
\label{fig:6}
\end{center}
\end{figure*}

Figure~\ref{fig:stats} reveals that for each type of topology, a
threshold value $\sigma_t$ exists above which global amplitude death  almost
certainly sets in, i.e., $p_s(\sigma)\approx 0$ for $\sigma>\sigma_t$ (see solid lines in Fig.~\ref{fig:stats}). Comparing the different topologies, it is interesting to note that $\sigma_t$ is
smaller,  about 0.15 for $N=100$, in the case of the small-world and the random network as compared to the regular network, where
$\sigma_t$ is about 0.2. Furthermore, in small-world and random
networks $\sigma_t$ is preceded by a steep sigmoidal transition, while
in the case of the regular network the transition is less steep and
characterized by a long tail making  it difficult to clearly define
$\sigma_t$. As discussed later in detail, the reason for this behavior
is that in regular networks often spiking subnetworks survive, which
is very unlikely for small world and random networks. 
Noteworthy is also that in the case of regular networks global amplitude death occurs later for larger networks, while in small-world
and random networks {\textit small} networks survive longer.

The topology of the network is even more critical when considering the
fraction of highly synchronized networks, i.e., the curves for $p_h$
(dashed lines in Fig.~\ref{fig:stats}).
 A particularly  interesting phenomenon is the non-monotonic behavior of
 $p_h$  in the regular ring networks (Fig. \ref{fig:stats}(a)
 dashed). While the fraction of highly synchronized networks rapidly
 drops for
 small $\sigma$, it rises again as the fraction of spiking networks
 falls. For large $\sigma$, $p_h$ and $p_s$ converge,
 i.e., only highly synchronized   networks survive for higher
 $\sigma$. This is certainly not the case for intermediate  $\sigma$, where
 most of the surviving networks are not highly-synchronized.
 This counterintuitive behavior of $p_h$ can be explained by the fact that for intermediate  $\sigma$
 the network splits into subnetworks as shown in Fig.~\ref{fig:6}:
 Panel (a) shows the spiking pattern of a regular ring for intermediate
 $\sigma$ ( $\sigma=0.1$), i.e., a value for which almost all networks are spiking
 in a highly asynchronous manner; Panel (b) depicts the corresponding
 Kuramoto order parameter $R$. Panels (c) and (d) show the same for a longer time series up to  $t=2000$. It can be clearly seen that after some transient time ($t
 \approx 500$), the network consists of two different subnetworks, at
 times separated by a patch of partial amplitude death ({\it subnetwork
   synchronization}). Simulations
 indicate that these subnetworks survive for infinitely long times and
 do not synchronize.
As a consequence the two networks have slightly different ISIs resulting in a beating behavior:  For a while the networks are almost synchronized, then their spiking times drift apart
 leading to slow oscillations in the Kuramoto order parameter. Thus, most
 of the times $R$ is much smaller than $0.99$ and the networks is not classified as highly synchronized.

If $\sigma$ is increased above approximately $0.13$, the system either
exhibits global amplitude death, or the spiking in one of the
subnetworks only dies out. The nodes in the surviving subnetwork remain highly synchronized, since they all interact and the distribution width is not yet too large. Thus almost all networks which do not undergo global amplitude death are
highly synchronized (recall that in the definition Eq.~(\ref{eq:Kuramoto}) of the order parameter only spiking nodes are considered). For even larger $\sigma$, global amplitude death can be observed in almost all network realizations.

In contrast, in small-world networks a different behavior can be
observed:  $p_s$ and $p_h$ do not coincide
for large $\sigma$ but the spiking networks survive longer than the
highly synchronized networks (see  Fig. \ref{fig:stats}(b)). The reason
is that in a small-world network  subnetworks do not easily
arise because the additional long range links connect the different
parts of the network. However, more and more disruptions occur for larger
$\sigma$. They decrease $R$ and are thus responsible for the
fact that less highly synchronized network persist.

Neither traveling disruptions nor partial amplitude death can be observed in a random network. Therefore, almost all surviving networks are highly synchronized, as shown in Fig. \ref{fig:stats}(c).

\section{Two discrete delay times}\label{sec:twodelays}
\begin{figure}
\begin{center}
\includegraphics[width=0.3\textwidth]{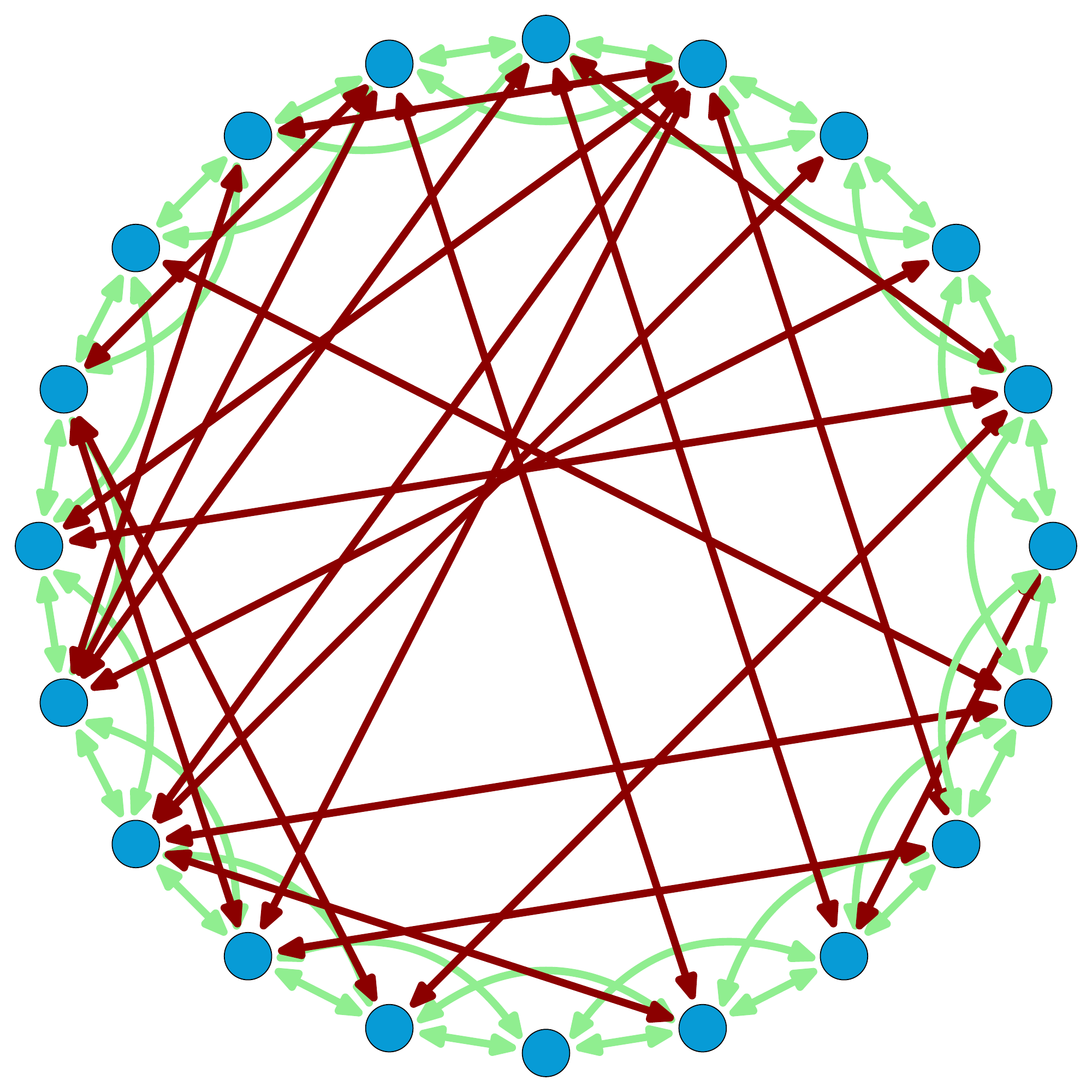}
\caption{Example of a small-world network, where inner connections
  (red (black) arrows) have a delay $\tau_{1}$, while the outer connections of
  the underlying regular ring (green (gray) arrows) are characterized by a
  delay $\tau_{2}$.}\label{fig:sw-inner-outer}
\end{center}
\end{figure}

\begin{figure*}
\begin{center}
 \begin{overpic}[width=1\textwidth]{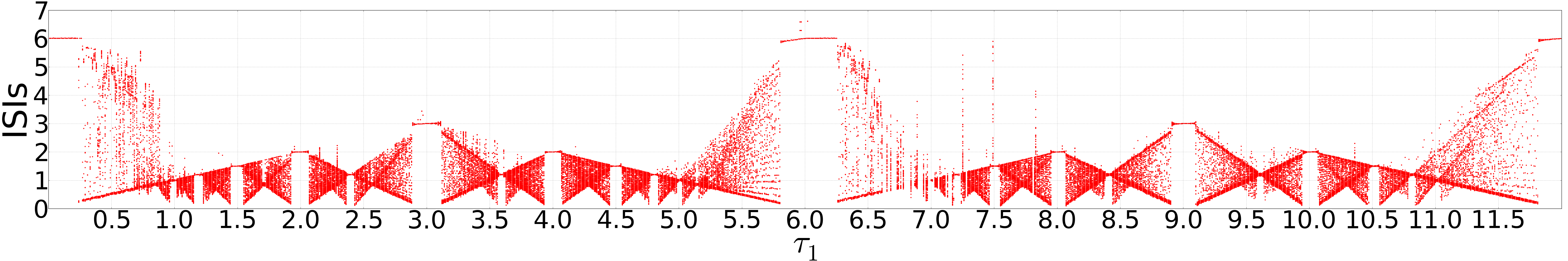}
  \put(4,14){(a)}
\end{overpic}
 \begin{overpic}[width=1\textwidth]{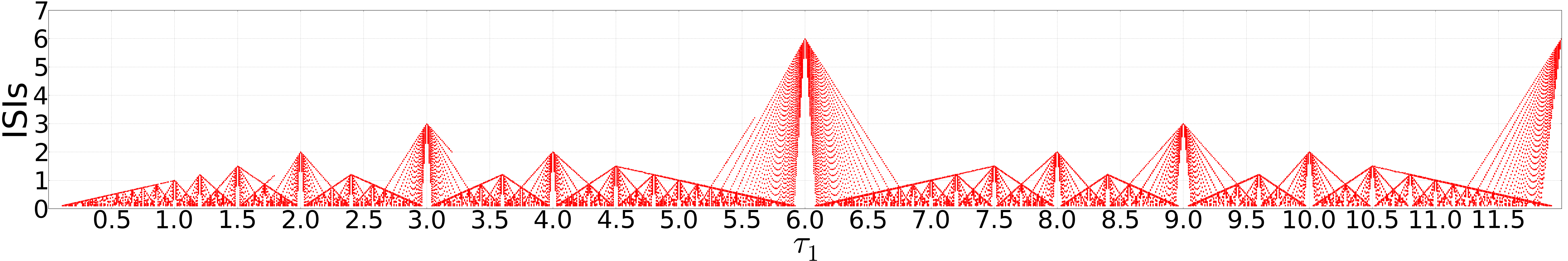}
  \put(4,14){(b)}
 \end{overpic}
\caption{Interspike intervals (ISIs) in a small-world network plotted versus the delay
  $\tau_{1}$ of the inner connections. (a) Simulations. (b)
  Mathematical reconstruction according to Eq. \eqref{math-sim}.
  $\tau_2=6$, $N=20$, $k=2$, $p=0.51$. Other parameters as in Fig.~\ref{fig:sync-commute}.
}\label{fig:sw-inner-outer-sim}
\end{center}
\end{figure*}
In Ref.~\cite{SCH08,PAN12} it was shown that already in a simple motif of
two coupled FitzHugh-Nagumo systems with two or three different delay times,
complex dynamics arise. In particular, resonance effects between the
different delay times proved to be crucial. Here we study the effect
of two discrete delay times in larger complex networks. We focus on
small-world networks (see Fig.~\ref{fig:sw-inner-outer} for a
schematic diagram) and separate the two parts of
the network in a meaningful way by choosing $\tau_2$ as the delay associated with the underlying regular network (green (gray) arrows), and $\tau_1$ as the delay time
of the additional random links (red (black) arrows).

Depending on the ratio between $\tau_{1}$ and $\tau_{2}$, different
spiking patterns emerge. We measure the ISIs (Interspike intervals)  in simulations while
gradually increasing $\tau_{1}$ for fixed $\tau_{2}=6$ as depicted in
Fig. \ref{fig:sw-inner-outer-sim}(a). Figure
\ref{fig:sw-inner-outer-sim}(b) shows a mathematical reconstruction of
the results obtained numerically in panel (a) based on the following
argument: Any spike in the network will eventually be fed into
the system again with a delay. Starting from the synchronous manifold,
i.e., $(u_1,v_1)=\ldots= (u_N,v_N) \equiv (u_s,v_s) $, spikes will first
reappear with a delay of either $\tau_{1}$ or $\tau_{2}$. Those spikes
again will be transmitted to other neurons with one of the two delays
meaning that eventually nearly all possible combinations of forwarding
a spike with either delay $\tau_{1}$ or $\tau_{2}$ will be observable
in the network. Thus, we  can obtain all possible spiking times from the expression:
\begin{align} T_{lk} = l \tau_{1} + k \tau_{2}
\end{align} \label{math-sim} with $l,k \in \mathbb{N}_0$. 

After sorting all possible $T_{lk}$ by size, the ISIs are given by the
difference between neighboring elements in the sorted list.  Note that
for the mathematical reconstruction in
Fig. \ref{fig:sw-inner-outer-sim}(b), ISIs $<0.1$ were discarded since
the spikes have a width of approximately 0.1 if they are close, i.e.,
the ISI is small. 

Coherent spiking, i.e., spiking with a constant ISI, is
observable as a result of resonance effects: For a case where the ratio of the multiple delay
times is given by 
\begin{align} n\tau_{1}=m\tau_{2}\label{eq:coherent_spiking}
\end{align} with $m,n \in \mathbb{N}$, coherent spiking with an ISI
equal to
\begin{align} \frac{\tau_{1}}{m}=\frac{\tau_{2}}{n
}\label{eq:isi_frac}
\end{align} is induced, where we choose the smallest possible $m$ and
$n$, i.e.,  $m$ and $n$ do not have common divisors.
A similar relation has been obtained
in Ref.~\cite{ZIG09} for a system of two time-discrete systems with
several unequal delays and in Ref.~\cite{PAN12} for a system of two
coupled FitzHugh-Nagumo systems with unequal coupling and
self-feedback delay times.
Note that $n$ and $m$ in Eq.~\eqref{eq:coherent_spiking} cannot be
chosen arbitrarily large: If $\tau_1/m$ or $\tau_2/n$ is smaller than
0.4 spikes run into each other and coherent spiking is not possible any longer.

\section{Bimodal delay distributions}
In this Section,  we discuss  a network characterized by a bimodal delay distribution. In such a network, the superposition of two
normal distributions with two mean delay times $\tau_{\mu}^{(1)}$ and
$\tau_{\mu}^{(2)}$ and two standard deviations $\sigma^{(1)}$ and
$\sigma^{(2)}$ determines the delay between nodes.

\subsubsection*{Peak distance of the distribution}
\begin{figure*}
\begin{center}
\includegraphics[width=1\textwidth]{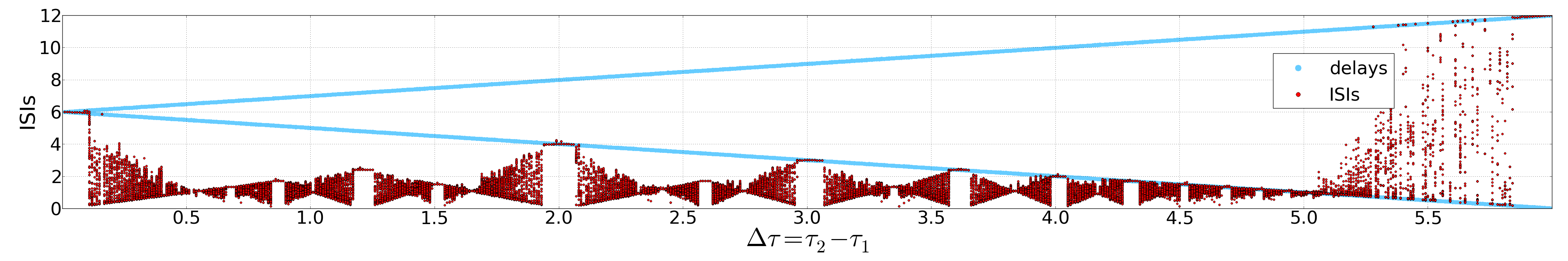}
\caption{Interspike intervals (ISIs) (red dots) and delays (blue dots) of a small-world network. $N=50$, $\sigma^{(1)}=\sigma^{(2)}=\sigma=0.01$, $p=0.51$. Other parameters as in Fig.~\ref{fig:sync-commute}.
}\label{fig:bi-mean}
\end{center}
\end{figure*}

To study the effects of different bimodal distributions, we start by changing the difference
$\tau_{\mu}^{(1)}-\tau_{\mu}^{(2)}$ between the two peaks of the
distribution, while keeping the standard deviations constant:
$\sigma^{(1)}=\sigma^{(2)}=0.01$. The result is shown in
Fig.~\ref{fig:bi-mean}, where the ISIs are depicted vs $\tau_{\mu}^{(1)}-\tau_{\mu}^{(2)}$.

Figure \ref{fig:bi-mean} shows that for distributions with small and medium width, i.e., $\sigma^{(1)}=\sigma^{(2)}<0.01$, the ISIs follow the condition of Eq. \eqref{eq:coherent_spiking} discussed for the case of two discrete delay times if we substitute $\tau_1$ and $\tau_2$ by the two peak positions, i.e.,  $\tau_1=\tau_{\mu}^{(1)}$ and $\tau_2=\tau_{\mu}^{(2)}$. The same pattern also emerges
for random networks (not shown here).

\subsubsection*{Width of the distribution}

\begin{figure*}
\begin{center}
\includegraphics[width=0.7\textwidth]{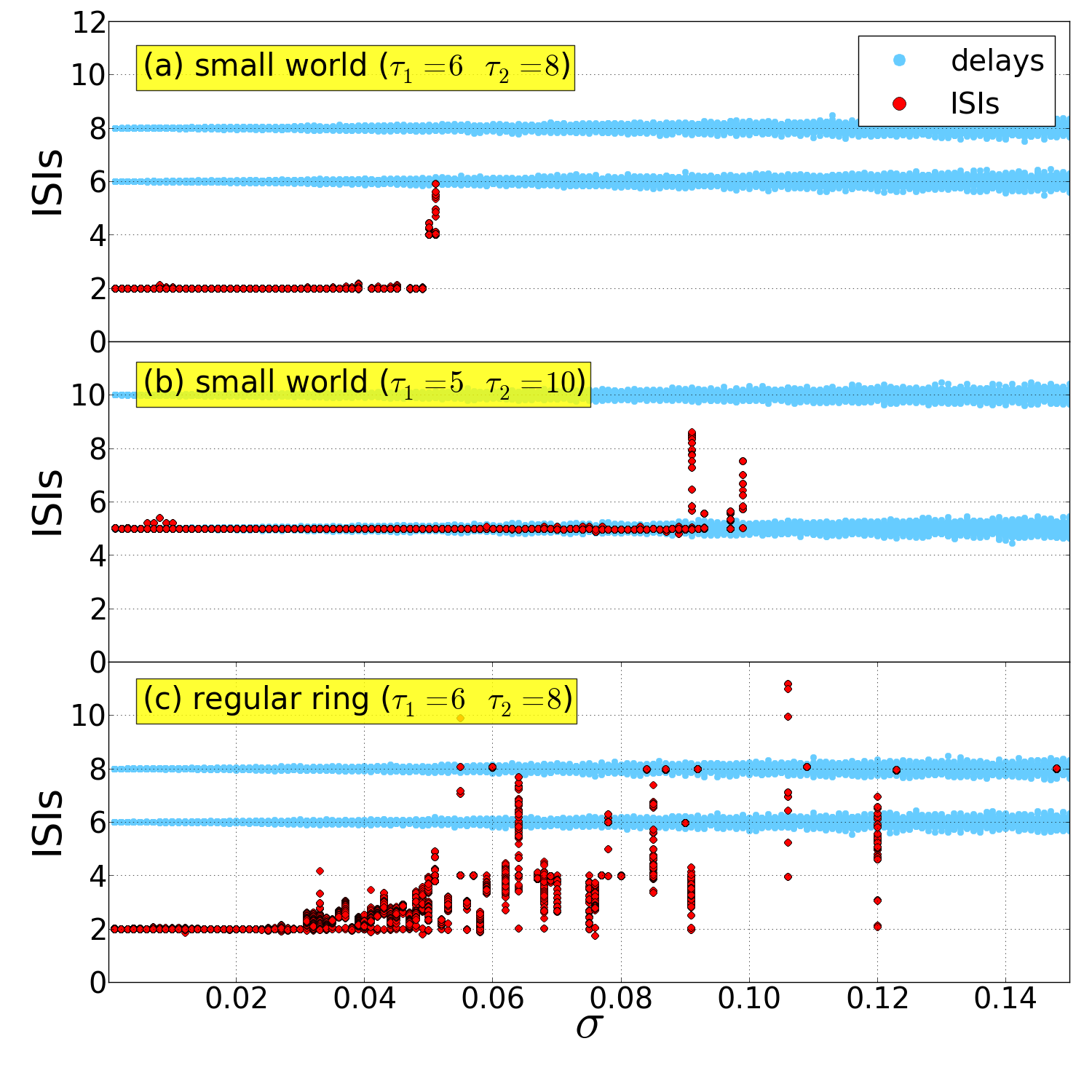}
\caption{Interspike intervals (ISIs) (red dots) and delay distributions (blue dots) vs $\sigma \equiv \sigma^{(1)}=\sigma^{(2)}$ for different combinations of mean delay times. (a), (b): small-world network ($k=2$, $p=0.51$), (c): regular ring network ($k=2$). Mean delay times are $\tau_1=6$ and
  $\tau_2=8$ for (a) and (c), and  $\tau_1=5$ and $\tau_2=10$ for
  (b). $N=50$.
Other parameters as in Fig.~\ref{fig:sync-commute}.
}\label{fig:bi-sw-sdev}
\end{center}
\end{figure*}

If the width of the two peaks becomes too large,
Eqs.~\eqref{eq:coherent_spiking}  and \eqref{eq:isi_frac} fail as
good descriptions for the spike times and ISIs.
Figure~\ref{fig:bi-sw-sdev} depicts the ISIs as function of the peak
widths $\sigma^{(1)}=\sigma^{(2)} \equiv\sigma$ for (a) a small world
network with $\tau_{\mu}^{(1)}=6$ and $\tau_{\mu}^{(2)}=8$, (b)  a small world
network with $\tau_{\mu}^{(1)}=5$ and $\tau_{\mu}^{(2)}=10$, and (c) a
regular ring with $\tau_{\mu}^{(1)}=6$ and $\tau_{\mu}^{(2)}=8$. For
small $\sigma$ ($\sigma<0.05$ in (a), $\sigma<0.09$ in (b), and
$\sigma<0.03$ in (c)) the ISIs can be found by evaluating
Eq.~\eqref{eq:isi_frac}:  For the combination $\tau_{\mu}^{(1)}=6$ and
$\tau_{\mu}^{(2)}=8$, we find coherent spiking with an ISI of 2, for
$\tau_{\mu}^{(1)}=5$ and $\tau_{\mu}^{(2)}=10$ the ISI is 5.

\begin{figure*}
\begin{center} 
\begin{overpic}[width=1\textwidth]{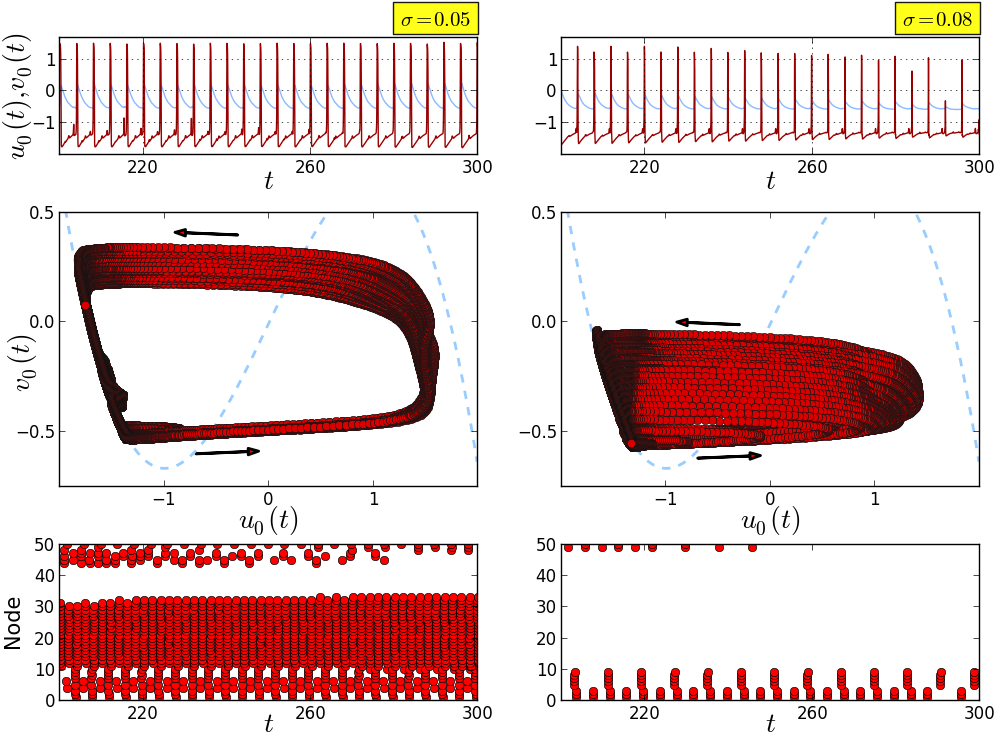}
\put(6.6,72.5){(a)}
\put(6.6,55){(b)}
\put(6.6,21.5){(c)}
\put(56,72.5){(d)}
\put(56,55){(e)}
\put(56,21.5){(f)}
\end{overpic}
\caption{ Dynamics for a bimodal delay distribution  in a regular
  ring with $k=2$ for different $\sigma$: (a)-(c) $\sigma=0.05$; (d)-(f)$ \sigma=0.08$. (a) and (d): time series of $u_0$ (dark red) and $v_0$ (light blue);  (b) and (e):  phase portraits of a single spiking
  node $(u_0, v_0)$: trajectory (red solid) and $u$-nullcline (blue dashed); (c)
  and (e): spiking patterns. Mean delay times $\tau_1=6$, $\tau_2=8$.
  Other parameters as in Fig.~\ref{fig:sync-commute}.
}\label{fig10}
\end{center}
\end{figure*}

As $\sigma$ increases, the coherent spiking breaks down.
Instead, networks can be observed  where different parts of the network spike with
different ISIs. This effect is particularly prominent in the case of a
ring network, since in such a network isolated subnetworks can easily 
arise, in which the delay distribution allows for persistent
spiking. Figure~\ref{fig10} shows exemplarily the dynamics in a
regular ring network for intermediate values of the distribution width ($\sigma=0.05$ in panels (a),(b),(c), $\sigma=0.08$ in panels (d),(e),(f)). Panels
(c) and (f) show the spiking patterns. For the lower $\sigma$ value,
(panel (c)), a part of the network (from about node 30 to node 45)
exhibits partial amplitude death, while the majority of the nodes keeps
spiking though in different subnetworks characterized by different ISIs. For the
higher  value of $\sigma$, only a small subset of spiking nodes persists.
The time series of node 0 in panel (a) and its phase portrait in (b) show that no longer all spikes have the same amplitude, but the amplitudes vary
slightly in an irregular fashion due to the coupling with other nodes with inhomogeneous delay times.

If $\sigma$ is further increased, global amplitude death sets in.

How robust the network is towards increasing the peak widths depends on the topology and the mean delay times. If $m$ and $n$
in Eq.~\eqref{eq:isi_frac} are large, coherent spiking is less robust,
since the probability that spikes overlap constructively after a time
$n\tau_{\mu}^{(1)}=m\tau_{\mu}^{(2)}$ decreases. For example, in the case
$\tau_{\mu}^{(1)}=6$ and $\tau_{\mu}^{(2)}=8$ $n=3$ and $m=2$, while
$\tau_{\mu}^{(1)}=5$ and $\tau_{\mu}^{(2)}=10$ yield the combination
$n=2$ and $m=1$, explaining why the synchronized spiking collapses in
Fig.~\ref{fig:bi-sw-sdev}(a)  for smaller $\sigma$ than in
Fig.~\ref{fig:bi-sw-sdev}(b). The size of the interval of $\sigma$, in
which asynchronous spiking takes place, depends on the topology. In regular rings
the interval is considerably  larger (see Fig.~\ref{fig:bi-sw-sdev}(c)) because subnetwork synchronization, as already observed in unimodal delay distributions, can occur. 
This subnetwork synchronization causes ISIs different than the ones predicted using
Eq.~\eqref{eq:coherent_spiking} while the spiking can still remain regular.
In random networks (not shown here), the interval shrinks to zero as
no regularity is left in the topology.

\section{Conclusion}
We have studied the effects of heterogeneous coupling delays in
complex networks of excitable elements. As a model we have used the
FitzHugh- Nagumo system which is generic for excitability of type II,
i.e., close to a Hopf bifurcation.  We have investigated two discrete
delay times as well as uni- and bimodal continuous distributions. As
topologies we have considered regular, small-world, and random
networks.

In case of unimodal distributions, we have found three different
dynamical scenarios: For narrow distributions the network fires in a
highly synchronized mode, because it behaves almost as expected for a
network with only one discrete delay time equal to the mean of the
distribution. Thus, such networks can be described in good
approximation by a discrete delay time which allows for further
analysis, for example, using the master stability function. If the
width of the distribution is increased, states might arise where the
network still fires but with reduced synchronicity. Whether such
states, for instance traveling disruption patterns, subnetwork
synchronization, or partial amplitude death, exist depends on the
topology since they require a certain degree of regularity found in
regular rings and to a lesser extent also in small-world networks. In
contrast, in random networks, i.e., in the absence of any regularity,
such dynamics is hardly found. If the width of the delay distribution
becomes too large, global amplitude death is induced.

Global amplitude death has first been observed by Reddy et. al. in
delay-coupled Stuart-Landau oscillators \cite{RED98}. They showed that an 
appropriate choice of the delay time changes the stability of the fixed point 
and yields global amplitude death.  In contrast, FitzHugh-Nagumo is an excitable system. 
Thus, delay in the coupling is needed to generate self-sustained oscillation with 
a period close to the delay time \cite{PAN12}. However, as we show, if the delay 
distribution becomes too large the delay cannot play this constructive role any longer.

The phenomena of partial amplitude death has been first investigated
by Atay in two coupled weakly nonlinear oscillators
\cite{ATA03a}. However, there partial amplitude death was induced by
heterogeneities in the nodes, while here we focus on delay
heterogeneities.

Masoller et. al. studied Gaussian and exponentially distributed delays
in networks of logistic maps \cite{MAS11a}. They also find synchronous
oscillating dynamics for narrow delay distributions and amplitude
death for wider distributions. However, the network topology plays a
different role; for logistic maps subnetwork synchronization is possible
in random networks, while we see this dynamical behavior only
in regular and (rarely) in small-world networks.

Furthermore, we investigated two discrete delay time and biomodial
distributions.  The dynamics in networks with two discrete delay times
is characterized by resonance effects, similar to the effects observed
in small network motifs \cite{SCH08,PAN12}, independent of
topology. If a resonance condition is fulfilled the network spikes
coherently with an interspike interval which is described by a simple
linear relation.

Bimodal distributions combine the features of the two cases discussed
above: They are characterized by two dominant mean delay times, but
with a distribution of the delays around these two peaks with some
widths. Hence, we observe dynamical patterns which we already
encountered in the two other cases. If the widths around the two mean
delays are small, the network behaves as in the case of two discrete
delay times and resonance effects play a major role. If the widths of
the distributions increases, we see dynamical scenarios already
present in the case of unimodal distributions with intermediate
width. In particular, in regular networks and small-world networks we
see that several subnetworks coexist which spike with different
interspike intervals. For large distribution widths, amplitude death 
is the only dynamical state which we have found.

In summary, networks with narrow delay distributions can be well
described by discrete delays, but as the width of the distribution
increases, topological features have to be taken into account as they
can give rise to more complex dynamics. In networks with broad
distributions spiking is not possible any longer and initial
excitations die out fast, leading to global amplitude death. This
behavior is similar to the case of only two coupled oscillators with
distributed delay in the link between them, where the regime of
amplitude death increases with the width of the delay kernel
\cite{ATA03,KYR11,KYR13}. The robustness of highly synchronous spiking
against heterogeneous delays is best for random networks, and worst
for large regular networks, and intermediate for small-world
networks. In contrast, large regular networks are more robust than
random or small-world networks with respect to avoiding global
amplitude death, since they can allow stable subnetwork spiking more
easily.

\subsubsection*{Acknowledgments.} This work was supported by the DFG in the framework of the SFB 910.

\begin{thebibliography}{40}

\bibitem{PIK01}
A.S. Pikovsky, M.G. Rosenblum, J.~Kurths, \emph{Synchronization, A Universal
  Concept in Nonlinear Sciences} (Cambridge University Press, Cambridge, 2001)

\bibitem{ROE97}
P.~Roelfsema, A.~Engel, P.~König, W.~Singer, Nature \textbf{385}, 157 (1997)

\bibitem{POE01}
K.~Poeck, W.~Hacke, \emph{Neurologie}, 11th~edn. (Springer, Heidelberg, 2001)

\bibitem{ROS05}
E.~Rossoni, Y.~Chen, M.~Ding, J.~Feng, Phys.~Rev.~E \textbf{71}, 061904 (~11)
  (2005)

\bibitem{SIN07}
W.~Singer, Scholarpedia \textbf{2}, 1657 (2007)

\bibitem{VIC08}
R.~Vicente, L.L. Gollo, C.R. Mirasso, I.~Fischer, P.~Gordon, Proc. Natl. Acad.
  Sci. U.S.A. \textbf{105}, 17157 (2008)

\bibitem{MAS08}
C.~Masoller, M.C. Torrent, J.~Garc\'\i{}a-Ojalvo, Phys. Rev. E \textbf{78},
  041907 (2008)

\bibitem{LEH11}
J.~Lehnert, T.~Dahms, P.~H{\"o}vel, E.~Sch{\"o}ll, Europhys. Lett. \textbf{96},
  60013 (2011)

\bibitem{KAN11a}
I.~Kanter, E.~Kopelowitz, R.~Vardi, M.~Zigzag, W.~Kinzel, M.~Abeles, D.~Cohen,
  Europhys.~Lett. \textbf{93}, 66001 (2011)

\bibitem{PER11a}
T.~P\'erez, G.C. Garcia, V.M. Eguiluz, R.~Vicente, G.~Pipa, C.~Mirasso, PLoS
  ONE \textbf{6}, e19900 (2011)

\bibitem{KEA12}
A.~Keane, T.~Dahms, J.~Lehnert, S.A. Suryanarayana, P.~H{\"o}vel,
  E.~Sch{\"o}ll, Eur. Phys. J.~B \textbf{85}, 407 (2012)

\bibitem{PEC98}
L.M. Pecora, T.L. Carroll, Phys. Rev. Lett. \textbf{80}, 2109 (1998)

\bibitem{KES07}
J.~Kestler, W.~Kinzel, I.~Kanter, Phys.~Rev.~E \textbf{76}, 035202 (~4) (2007)

\bibitem{SOR07}
F.~Sorrentino, E.~Ott, Phys. Rev.~E \textbf{76}, 056114 (~10) (2007)

\bibitem{DAH12}
T.~Dahms, J.~Lehnert, E.~Sch{\"o}ll, Phys. Rev.~E \textbf{86}, 016202 (2012)

\bibitem{KOC99}
C.~Koch, \emph{Biophysics of Computation: Information Processing in Single
  Neurons} (Oxford University Press, New York, 1999)

\bibitem{FIT61}
R.~FitzHugh, Biophys. J. \textbf{1}, 445 (1961)

\bibitem{NAG62}
J.~Nagumo, S.~Arimoto, S.~Yoshizawa., Proc. IRE \textbf{50}, 2061 (1962)

\bibitem{LIN04}
B.~Lindner, J.~Garc{\'i}a-Ojalvo, A.~Neiman, L.~Schimansky-Geier, Phys.~Rep.
  \textbf{392}, 321 (2004)

\bibitem{HEI10}
M.~Heinrich, T.~Dahms, V.~Flunkert, S.W. Teitsworth, E.~Sch{\"o}ll,
  New~J.~Phys. \textbf{12}, 113030 (2010)

\bibitem{MUR93}
J.D. Murray, \emph{Mathematical Biology}, Vol.~19 of \emph{Biomathematics
  Texts}, 2nd~edn. (Springer, Berlin Heidelberg, 1993)

\bibitem{IZH00a}
E.M. Izhikevich, Int. J. Bifurc. Chaos \textbf{10}, 1171 (2000)

\bibitem{WAT98}
D.J. Watts, S.H. Strogatz, Nature \textbf{393}, 440 (1998)

\bibitem{MON99}
R.~Monasson, Eur. Phys. J. B \textbf{12}, 555 (1999)

\bibitem{NEW99b}
M.E.J. Newman, D.J. Watts, Phys. Lett. A \textbf{263}, 341 (1999)

\bibitem{RAP57}
A.~Rapoport, Bull.~Math.~Biol. \textbf{19}, 257 (1957)

\bibitem{SOL51}
R.~Solomonoff, A.~Rapoport, Bull.~Math.~Biol. \textbf{13}, 107 (1951)

\bibitem{ERD59}
P.~Erd\H{o}s, A.~R\'{e}nyi, Publ. Math. Debrecen \textbf{6}, 290 (1959)

\bibitem{ERD60}
P.~Erd\H{o}s, A.~R\'{e}nyi, Publ. Math. Inst. Hung. Acad. Sci \textbf{5}, 17
  (1960)

\bibitem{KUR84}
Y.~Kuramoto, \emph{Chemical Oscillations, Waves and Turbulence}
  (Springer-Verlag, Berlin, 1984)

\bibitem{ROS01}
M.G. Rosenblum, A.S. Pikovsky, J.~Kurths, C.~Sch{\"a}fer, P.A. Tass,
  \emph{{Phase synchronization: from theory to data analysis}} (Elsevier
  Science, Amsterdam, 2001), Vol.~4 of \emph{{Handbook of Biological Physics}},
  chap.~9, pp. 279--321, 1st~edn.

\bibitem{SCH08}
E.~Sch{\"o}ll, G.~Hiller, P.~H{\"o}vel, M.A. Dahlem, Phil. Trans.~R. Soc.~A
  \textbf{367}, 1079 (2009)

\bibitem{PAN12}
A.~Panchuk, D.P. Rosin, P.~H{\"o}vel, E.~Sch{\"o}ll, Int. J.~Bif. Chaos
  \textbf{23}, 1330039 (2013)

\bibitem{ZIG09}
M.~Zigzag, M.~Butkovski, A.~Englert, W.~Kinzel, I.~Kanter, Europhys. Lett.
  \textbf{85}, 60005 (2009)

\bibitem{RED98}
D.V.R. Reddy, A.~Sen, G.L. Johnston, Phys. Rev. Lett. \textbf{80}, 5109 (1998)

\bibitem{ATA03a}
F.M. Atay, Physica D \textbf{183}, 1 (2003)

\bibitem{MAS11a}
C.~Masoller, F.M. Atay, Eur. Phys. J. D \textbf{62}, 119 (2011)

\bibitem{ATA03}
F.M. Atay, Phys.~Rev.~Lett. \textbf{91}, 094101 (2003)

\bibitem{KYR11}
Y.N. Kyrychko, K.B. Blyuss, E.~Sch{\"o}ll, Eur. Phys.~J.~B \textbf{84}, 307
  (2011)

\bibitem{KYR13}
Y.N. Kyrychko, K.B. Blyuss, E.~Sch{\"o}ll, Phil. Trans.~R. Soc.~A \textbf{371},
  20120466 (2013)

\end{thebibliography}

\end{document}